\documentstyle[12pt,epsfig]{article}
\newcommand{\nd}{\noindent}
\newcommand{\beq}{\begin{equation}}
\newcommand{\eeq}{\end{equation}}
\newcommand{\barr}{\begin{eqnarray}}
\newcommand{\earr}{\end{eqnarray}}
\newcommand{\ba}{\begin{array}}
\newcommand{\ea}{\end{array}}

\newcommand{\bfp}{\mbox{\boldmath $p$}}
\newcommand{\bfP}{\mbox{\boldmath $P$}}
\newcommand{\bfk}{\mbox{\boldmath $k$}}

\newcommand{\bfl}{\mbox{\boldmath $l$}}

\newcommand{\qup}{q^\uparrow}

\newcommand{\qdown}{q^\downarrow}
\newcommand{\NP}[1]{{\it Nucl.\ Phys.}\ {\bf #1}}
\newcommand{\ZP}[1]{{\it Z.\ Phys.}\ {\bf #1}}
\newcommand{\PL}[1]{{\it Phys.\ Lett.}\ {\bf #1}}
\newcommand{\PR}[1]{{\it Phys.\ Rev.}\ {\bf #1}}
\newcommand{\PRL}[1]{{\it Phys.\ Rev.\ Lett.}\ {\bf #1}}
\newcommand{\IJMP}[1]{{\it Int.\ J.\ Mod.\ Phys.}\ {\bf #1}}

\def\lsim{\mathrel{\rlap{\lower4pt\hbox{\hskip1pt$\sim$}}\raise1pt\hbox{$<$}}}
\def\gsim{\mathrel{\rlap{\lower4pt\hbox{\hskip1pt$\sim$}}\raise1pt\hbox{$>$}}}
\def\nostrocostruttino#1\over#2{\mathrel{\mathop{\kern 0pt \rlap
{\hbox{$#1$}}} \hbox{\kern-.135em $#2$}}}

\begin{document}
\begin{flushright}
INFNCA-TH0001 \\
hep-ph/0001307 \\
\end{flushright}
\vskip 1.5cm
\begin{center}
{\bf $\Lambda$ and $\bar\Lambda$ polarization in polarized DIS}\\
\vskip 0.8cm
{\sf M.\ Anselmino$^1$, M.\ Boglione$^2$ and F.\ Murgia$^3$}
\vskip 0.5cm
{$^1$ Dipartimento di Fisica Teorica, Universit\`a di Torino and \\
      INFN, Sezione di Torino, Via P. Giuria 1, I-10125 Torino, Italy\\
\vskip 0.5cm
$^2$  Dept. of Physics and Astronomy, Vrije Universiteit Amsterdam, \\
De Boelelaan 1081, 1081 HV Amsterdam, The Netherlands \\ 
\vskip 0.5cm
$^3$  Istituto Nazionale di Fisica Nucleare, Sezione di Cagliari\\
      and Dipartimento di Fisica, Universit\`a di Cagliari\\
      C.P. 170, I-09042 Monserrato (CA), Italy} \\
\end{center}
\vskip 1.5cm
\noindent
{\bf Abstract:} \\ 
We consider the polarization of $\Lambda + \bar\Lambda$ baryons 
produced in polarized Deep Inelastic Scattering at leading order, with 
various spin configurations: longitudinally polarized leptons and 
unpolarized nucleon; unpolarized leptons and longitudinally or transversely 
polarized nucleons; longitudinally polarized leptons and nucleons. We show 
how the different results in the different cases are related to different 
aspects of the elementary dynamics and to the spin properties of the 
distribution and fragmentation functions and show how a combined analysis 
might give useful information. We give numerical results according 
to several sets of polarized fragmentation functions recently proposed.
\newpage
\pagestyle{plain}
\setcounter{page}{1}
\nd
{\bf 1. Introduction}
\vskip 6pt
$\Lambda$ baryons produced in high energy interactions and resulting 
from quark fragmentation allow a unique test of spin transfer 
from partons to hadrons: the $\Lambda$ polarization is easily
measurable by looking at the angular distribution of the 
$\Lambda \to p \pi$ decay (in the $\Lambda$ helicity rest frame) and
the fragmenting parton polarization is determined by the elementary
Standard Model interactions, provided one knows the initial parton
spin state. In this respect $\Lambda$'s produced in lepton induced
processes are particularly interesting and indeed several 
papers on this subject have recently been published or submitted
to e-Print archives \cite{jaf}-\cite{mssy}.

We perform here a detailed analysis of the polarization of $\Lambda$'s 
produced in polarized DIS; a general discussion of the helicity density 
matrix of hadrons produced in polarized lepton-nucleon interactions, 
at leading order, can be found in Refs. \cite{noi} and \cite{jaf}. 
It can be written as:
\barr
\rho_{\lambda^{\,}_h,\lambda^\prime_h}^{(s,S)}(h) \>
\frac{d\sigma^{\ell,s + N,S \to \ell + h + X}} {dx \> dy \> dz} &=& 
\sum_{q; \lambda^{\,}_{\ell}, \lambda^{\,}_q, \lambda^\prime_q}
\frac {1}{16 \pi x s} \> \times \label{gen} \\  
& & \!\!\!\! \rho^{\ell,s}_{\lambda^{\,}_{\ell}, \lambda^{\,}_{\ell}} \,
\rho_{\lambda^{\,}_q, \lambda^{\prime}_q}^{q/N,S} \, f_{q/N}(x) \,
\hat M^q_{\lambda^{\,}_{\ell}, \lambda^{\,}_q; 
\lambda^{\,}_{\ell}, \lambda^{\,}_q} \, 
\hat M^{q*}_{\lambda^{\,}_{\ell}, \lambda^{\prime}_q; 
\lambda^{\,}_{\ell}, \lambda^{\prime}_q} \,
D_{\lambda^{\,}_h, \lambda^{\prime}_h}^{\lambda^{\,}_q,\lambda^{\prime}_q}(z) 
\>, \nonumber
\earr
where $x$ and $y$ are the usual DIS variables, $x = Q^2/2p \cdot q$,
$y = Q^2/xs$ and, neglecting hadron masses, $z = p_h \cdot p/ p \cdot q$,
where $p$, $q$ and $p_h$ are, respectively, the nucleon, virtual
photon and final hadron four-momenta. 
$\rho^{\ell,s}$ is the helicity density matrix of the initial 
lepton with spin $s$, $f_{q/N}(x)$ is the number density of unpolarized 
quarks $q$ with momentum fraction $x$ inside an unpolarized nucleon and 
$\rho^{q/N,S}$ is the helicity density matrix of quark $q$ inside the 
polarized nucleon $N$ with spin $S$. 
The $\hat M^q_{\lambda^{\,}_{\ell}, \lambda^{\,}_q; \lambda^{\,}_{\ell}, 
\lambda^{\,}_q}$'s are the helicity amplitudes for the elementary process 
$\ell q \to \ell q$. The final lepton spin is not observed and helicity 
conservation of perturbative QCD and QED has already been taken into account
in the above equation: as a consequence only the diagonal elements of 
$\rho^{\ell,s}$ contribute to $\rho(h)$ and non diagonal elements, present in 
case of transversely polarized leptons, do not contribute.
$D^{\lambda^{\,}_q,\lambda^{\prime}_q}_{\lambda^{\,}_h,\lambda^{\prime}_h}(z)$ 
is a generalized fragmentation function \cite{noi} related to the 
usual unpolarized fragmentation function $D_{h/q}(z)$, {\it i.e.} 
the density number of hadrons $h$ resulting from the fragmentation of 
an unpolarized quark $q$ and carrying a fraction $z$ of its momentum, by
\begin{equation}
D_{h/q}(z) = {1\over 2} \sum_{\lambda^{\,}_q, \lambda^{\,}_h} 
D^{\lambda^{\,}_q,\lambda^{\,}_q}_{\lambda^{\,}_h,\lambda^{\,}_h}(z) 
= {1\over 2} \sum_{\lambda^{\,}_q, \lambda^{\,}_h} 
D_{h_{\lambda^{\,}_h}/q_{\lambda^{\,}_q}}(z) \,,
\label{fr}
\end{equation}
where $D^{\lambda^{\,}_q,\lambda^{\,}_q}_{\lambda^{\,}_h,
\lambda^{\,}_h}(z) \equiv D_{h_{\lambda^{\,}_h}/q_{\lambda^{\,}_q}}(z)$ is
a polarized fragmentation function, {\it i.e.} the density number of 
hadrons $h$ with helicity $\lambda^{\,}_h$ resulting from the fragmentation 
of a quark $q$  with helicity $\lambda^{\,}_q$. Angular momentum 
conservation and collinear configurations imply for the generalized 
fragmentation functions \cite{noi}:
\beq
D^{\lambda^{\,}_q,\lambda^{\prime}_q}_{\lambda^{\,}_h,\lambda^{\prime}_h}
= 0 \quad\quad {\mbox{\rm when}} \quad\quad
\lambda^{\,}_q - \lambda^{\prime}_q \not= \lambda^{\,}_h-\lambda^{\prime}_h \,.
\label{for}
\eeq
The elementary amplitudes $\hat M^q$ are normalized so that the elementary 
unpolarized cross-section, $\ell q \to \ell q$, is:
\beq 
{d\hat\sigma^q \over dy} \equiv e_q^2 \> {d\hat\sigma \over dy} = 
{1\over 16\pi x s} \>
{1 \over 4} \, \sum_{\lambda^{\,}_{\ell}, \lambda^{\,}_q}
|\hat M^q_{\lambda^{\,}_{\ell}, \lambda^{\,}_q;
\lambda^{\,}_{\ell}, \lambda^{\,}_q}|^2 =
e_q^2 \> {2 \pi \alpha^2 \over Q^2} \> {1 + (1-y)^2 \over y} \> \cdot
\label{norm}
\eeq

Notice that there are only two independent helicity amplitudes:
\barr
\hat M^q_{++;++} = \hat M^q_{--;--}
&\equiv& e_q \> \hat M_{++;++} = e_q \> 8 \pi \alpha 
{1 \over y} \nonumber \\
\hat M^q_{+-;+-} = \hat M^q_{-+;-+}
&\equiv& e_q \> \hat M_{+-;+-} = e_q \> 8 \pi \alpha 
\left[{1 \over y} - 1 \right] \label{helamp}
\earr 
so that
\beq 
{d\hat\sigma^q \over dy} = {1 \over 2}\> \left[ {d\hat\sigma^{++}_q \over dy}
+ {d\hat\sigma^{+-}_q \over dy} \right] \label{dsun}
\eeq
with
\beq 
{d\hat\sigma^{++}_q \over dy} \equiv e_q^2 \> {d\hat\sigma^{++} \over dy}
= {e_q^2\over 16\pi x s} \> |\hat M_{++;++}|^2 =
e_q^2 \> {4 \pi \alpha^2 \over Q^2} \> {1 \over y} \label{ds++}
\eeq
and
\beq 
{d\hat\sigma^{+-}_q \over dy} \equiv e_q^2 \> {d\hat\sigma^{+-} \over dy}
= {e_q^2\over 16\pi x s} \> |\hat M_{+-;+-}|^2 =
e_q^2 \> {4 \pi \alpha^2 \over Q^2} \> {(1-y)^2 \over y} \> \cdot \label{ds+-}
\eeq
 
Finally, the cross-section appearing in the l.h.s. of Eq. (\ref{gen}),
which gives the correct normalization to $\rho(h)$, Tr$\rho = 1$,  
can be written, using Eq. (\ref{for}), as
\beq
\frac{d\sigma^{\ell,s + N,S \to \ell + h + X}} {dx \> dy \> dz} = 
\sum_{q; \lambda^{\,}_{\ell}, \lambda^{\,}_q}
\frac {1}{16 \pi x s}\>
\rho^{\ell,s}_{\lambda^{\,}_{\ell}, \lambda^{\,}_{\ell}} \,
\rho_{\lambda^{\,}_q, \lambda^{\,}_q}^{q/N,S} \, f_{q/N}(x) \,
|\hat M^q_{\lambda^{\,}_{\ell}, \lambda^{\,}_q; 
\lambda^{\,}_{\ell}, \lambda^{\,}_q}(y)|^2 D_{h/q}(z) \>.
\label{dspol}
\eeq

Eqs. (\ref{gen})-(\ref{dspol}) hold within QCD factorization
theorem at leading twist and leading order in the coupling constants;
the intrinsic $\bfk_\perp$ of the partons have been integrated over and 
collinear configurations dominate both the distribution and 
the fragmentation functions. For simplicity of notations we have not 
indicated the $Q^2$ scale dependences in $f$ and $D$. 

We shall use such equations for spin 1/2 $\Lambda$ baryons produced starting
from several particular initial spin configurations; we will discuss how the 
measurable components of the $\Lambda$ polarization vector depend on different
combinations of distribution functions, elementary dynamics and fragmentation
functions: each of these terms predominantly depends on a single variable, 
respectively $x$, $y$ and $z$, and a careful analysis of different
situations can yield precious information. Although $\Lambda$ production
in polarized DIS has been recently discussed in several papers, most of them 
only consider some initial spin configurations and specific models for 
fragmentation functions. Our analysis is more comprehensive and somewhat 
more general, emphasizing the physical meaning of possible measurements,
and allowing also to obtain some general relationships between
different polarization values.     

\vskip 18pt
\nd
{\bf 2. Polarization vector of spin 1/2 baryons}
\vskip 6pt
We fix our spin notations in the $\ell-p$ center of mass frame:
the lepton moves with four-momentum $l$ along the $z$-axis
and the proton moves with four-momentum $p$ in the opposite 
direction; we choose $xz$ as the lepton-hadron production plane, 
with the $y$-axis parallel to $\bfl \times \bfp_h$. We denote 
by $S_L$ the (longitudinal) nucleon spin oriented along the 
$z$-axis, by $S_S$ the (sideway) spin oriented along the $x$-axis and  
by $S_N$ the (normal) spin oriented along the $y$-axis. Notice that
$+ S_L$ corresponds to a $-$ helicity proton and $-S_L$ to a $+$ helicity
one; we will only consider longitudinally polarized leptons with
spins $\pm s_L$ which correspond respectively to $\pm$ helicities.

From Eqs. (\ref{gen})-(\ref{dspol}) one obtains the explicit expression 
for the components of the helicity density matrix of a spin 1/2 baryon. 
These are related to the three components of the baryon polarization vector, 
as measured in its helicity rest frame by $P_i(B) =$ Tr$[(\sigma^i\rho(B)]$
($i = x,y,z$); details can be found in Ref. \cite{noi}. We denote by
$(s, S)$ [or $(h, H)$] the (lepton, nucleon) spins [or helicities], 
0 stands for unpolarized particle; one finds, for a spin 1/2 hadron $B$:   

\barr
P_z^{(0, -S_L)}(B; x,y,z) &=& P_z^{(0,+)}(B; x,y,z) = 
\frac{\sum_q \Delta q \> d\hat\sigma^q \>
\Delta D_{B/q}} {\sum_q q \> d\hat\sigma^q \> D_{B/q}} \nonumber \\
&=& \frac{\sum_q e_q^2 \> \Delta q(x) \> \Delta D_{B/q}(z)} 
{\sum_q e_q^2 \> q(x) \> D_{B/q}(z)} \label{0-SL} \\ 
P_z^{(s_L, 0)}(B; x,y,z) &=& P_z^{(+,0)}(B; x,y,z) = \frac
{\sum_q (1/2)\,q \> [d\hat\sigma_q^{++} - d\hat\sigma_q^{+-}]\> 
\Delta D_{B/q}} {\sum_q q \> d\hat\sigma^q \> D_{B/q}} \nonumber \\ 
&=& \frac{\sum_q e_q^2 \> q(x) \> \Delta D_{B/q}(z)} 
{\sum_q e_q^2 \> q(x) \> D_{B/q}(z)} \> \hat A_{LL}(y) \label{sL0} \\
P_y^{(0, S_N)}(B; x,y,z) &=& - P_x^{(0, S_S)}(B; x,y,z) =
\frac {\sum_q \Delta_T q \> [d\hat\sigma_q^{\uparrow \to \uparrow} - 
d\hat\sigma_q^{\uparrow \to \downarrow}]\> \Delta_T D_{B/q}} 
{\sum_q q \> d\hat\sigma^q \> D_{B/q}} \nonumber \\ 
&=& \frac{\sum_q e_q^2 \> \Delta_T q(x) \> \Delta_T D_{B/q}(z)} 
{\sum_q e_q^2 \> q(x) \> D_{B/q}(z)} \> \hat D_{NN}(y) \label{0SN} \\
P_z^{(s_L, -S_L)}(B; x,y,z) &=& P_z^{(+,+)}(B; x,y,z) =
\frac{\sum_q [q_+ \> d\hat\sigma_q^{++} - q_- \> d\hat\sigma_q^{+-}] \>
\Delta D_{B/q}} {\sum_q [q_+ \> d\hat\sigma_q^{++} 
+ q_- \> d\hat\sigma_q^{+-}] \> D_{B/q}} \nonumber \\
&=& \frac{\sum_q e_q^2 \> [q(x) \hat A_{LL}(y) + \Delta q(x)] 
\> \Delta D_{B/q}(z)} 
{\sum_q e_q^2 \> [q(x) + \hat A_{LL}(y) \Delta q(x)] \> D_{B/q}(z)} 
\label{sL-SL} \\
P_z^{(s_L, +S_L)}(B; x,y,z) &=& P_z^{(+,-)}(B; x,y,z) =
\frac{\sum_q [q_- \> d\hat\sigma_q^{++} - q_+ \> d\hat\sigma_q^{+-}] \>
\Delta D_{B/q}} {\sum_q [q_- \> d\hat\sigma_q^{++} 
+ q_+ \> d\hat\sigma_q^{+-}] \> D_{B/q}} \nonumber \\
&=& \frac{\sum_q e_q^2 \> [q(x) \hat A_{LL}(y) - \Delta q(x)] 
\> \Delta D_{B/q}(z)} 
{\sum_q e_q^2 \> [q(x) - \hat A_{LL}(y) \Delta q(x)] \> D_{B/q}(z)} 
\label{sL+SL} 
\earr
where $d\hat\sigma_q$ stands for $d\hat\sigma_q/dy$.

Some comments are in order.
\begin{itemize}
\item
The above results are known \cite{jaf, ekk, am}; we have rederived
and grouped them here for convenience and further discussion. Notice that 
all other spin configurations -- at leading order -- either give the same 
results or no polarization. We remind that $q_\lambda = f_{q_\lambda/p_+}$ 
is the number density of quarks with helicity $\lambda$ inside a + helicity 
proton, $q(x) = q_+(x) + q_-(x)$ and $\Delta q(x) = q_+(x) - q_-(x)$. 
Similarly $\Delta D_{B/q} = D_{B_+/q_+} - D_{B_-/q_+}$;
$\Delta_Tq$ and $\Delta_T D_{B/q}$ are respectively the analogue of 
$\Delta q$ and $\Delta D_{B/q}$ for transverse spins.   

\item
The longitudinal $B$ polarization induced by a longitudinal
nucleon polarization, Eq. (\ref{0-SL}), {\it does not} depend on the 
elementary dynamics, but only on the quark spin distribution and
fragmentation properties. Neglecting the QCD $Q^2$-evolution, 
$P_z^{(0,-S_L)}$ does not depend on the DIS variable $y$, but only on $x$ 
and $z$.  

\item
The longitudinal $B$ polarization resulting from the scattering of
longitudinally polarized leptons off unpolarized nucleons depends on
the unpolarized distribution functions, the polarized fragmentation 
functions and the elementary dynamics, through the double spin asymmetry
for the $\ell q \to \ell q$ process [see Eqs. (\ref{dsun})-(\ref{ds+-})]:
\beq
\hat A_{LL} (y) = \frac 
{d\hat\sigma^{++}_q - d\hat\sigma^{+-}_q} 
{d\hat\sigma^{++}_q + d\hat\sigma^{+-}_q}
= \frac {d\hat\sigma^{++}_q - d\hat\sigma^{+-}_q}
{2 \, d\hat\sigma^q} = \frac {y(2-y)}{1+(1-y)^2} \> \cdot
\label{all}
\eeq

Notice that $\hat A_{LL}$ grows with $y$ from 0 (at $y=0$) to 1 (at $y=1$),
so that $P_z^{(s_L,0)}$ is an {\it increasing} function of $y$, starting 
from 0 at $y=0$.
\item
The transverse $B$ polarization induced by a transverse 
nucleon polarization, Eq. (\ref{0SN}), depends on the quark transverse spin 
distribution and fragmentation properties and on the elementary dynamics, 
through the double transverse spin asymmetry for the 
$\ell q \to \ell q$ process:
\barr
\hat D_{NN} (y) &=& \frac 
{d\hat\sigma^{\ell \qup \to \ell \qup} -
 d\hat\sigma^{\ell \qup \to \ell \qdown}}
{d\hat\sigma^{\ell \qup \to \ell \qup} +
 d\hat\sigma^{\ell \qup \to \ell \qdown}} \nonumber \\
&=& \frac {d\hat\sigma^{\ell \qup \to \ell \qup} -
 d\hat\sigma^{\ell \qup \to \ell \qdown}}
{d\hat\sigma^q} = \frac{2(1-y)}{1+(1-y)^2} \label{dnn}
\earr
where $\uparrow = S_N$ and $\downarrow = -S_N$.
 
Contrary to $\hat A_{LL}$, $\hat D_{NN}$ decreases with $y$,
with $\hat D_{NN}=1$ at $y=0$ and $\hat D_{NN}=0$ at $y=1$; thus,
$P_y^{(0,S_N)}$ is a {\it decreasing} function of $y$, reaching 0 at $y=1$.
An experimental confirmation of the  opposite $y$-dependences 
of $P_z^{(s_L,0)}$ and $P_y^{(0,S_N)}$ would supply a new, subtle 
and important test of the factorization scheme of Eq. (\ref{gen}).  
$\hat D_{NN}$ is the transverse polarization of the final quark generated 
by an initial transversely polarized ($\uparrow$) quark in the
$\ell q \to \ell q$ process, and it is usually referred to as the
depolarization factor. 
 
\item
When both the lepton and the nucleon are longitudinally polarized
the $B$ resulting polarization depends on yet a different combination
of polarized quark distribution functions, fragmentation functions and
elementary dynamics. It is then clear why a combined study of $P(B)$
in different cases could yield unique information. 
    
\item
Finally, we notice that, in general, $P_i^{(s,S)} = -P_i^{(-s,-S)}$.
\end{itemize}

\vskip 18pt
\nd
{\bf 3. Polarization vector of $\Lambda$ baryons}
\vskip 6pt

We now consider the particular case of $\Lambda$ baryons and discuss
possible ways of extracting information from combined measurements
of $P_i(\Lambda)$; as we said the $\Lambda$ polarization vector
can be measured by looking at the proton angular distribution as resulting
from $\Lambda \to \pi p$ decay in the $\Lambda$ helicity rest frame, that 
is the frame obtained by rotating the $\ell-p$ c.m. frame around the $y$-axis
so that the new $z_h$-axis is parallel to the $\Lambda$ direction, and 
then boosting along $z_h$ with the same speed as the $\Lambda$:
\barr
W(\theta_p, \phi_p) &=& {1 \over 4\pi} \left[ 1 + \alpha (P_z \cos\theta_p  
+ P_x \sin\theta_p\cos\phi_p + P_y \sin\theta_p\sin\phi_p) \right] \nonumber \\
&=& {1 \over 4\pi} \left[ 1 + \alpha \bfP \cdot \hat{\bfp} \right] \label{decl}
\earr
where $\alpha = 0.642 \pm 0.013$.

We follow Ref. \cite{vog} and assume for the unpolarized fragmentation 
functions:
\beq 
D_{\Lambda/u} = D_{\Lambda/d} = D_{\Lambda/s} =
D_{\Lambda/\bar u} = D_{\Lambda/\bar d} = D_{\Lambda/\bar s}
\equiv D_{\Lambda/q} \label{dl}
\eeq
where $\Lambda$ means $\Lambda^0 + \bar \Lambda^0$. 

The heavy quark and gluon unpolarized fragmentation functions play a 
negligible role for $z \gsim 0.3$ \cite{vog} and we neglect them 
here: we have actually checked that our results, when comparable, are almost 
indistinguishable from those of Ref. \cite{vog} where also heavy quark
and gluon contributions to the unpolarized cross-sections are taken into 
account. 

Similarly, we follow Ref. \cite{vog} for the polarized fragmentation 
functions:
%
\beq
\Delta D_{\Lambda/u}(z, Q^2_0) = \Delta D_{\Lambda/d}(z, Q^2_0) = 
N_u \, \Delta D_{\Lambda/s}(z, Q^2_0) \>. \label{ddq}
\eeq
Eqs. (\ref{ddq}) holds also for light antiquarks and it remains valid 
through QCD $Q^2$-evolution; heavy quark contributions are neglected.


Using Eq. (\ref{dl}) and (\ref{ddq}) into Eqs. (\ref{0-SL})-(\ref{sL0}) and
(\ref{sL-SL})-(\ref{sL+SL}) gives:
\barr
P_z^{(0,+)}(\Lambda; x,y,z) &=& \frac
{\Delta Q'(x)}{Q(x)} \> 
\frac{\Delta D_{\Lambda/s}(z)}{D_{\Lambda/q}(z)} \label{0-SLL} \\ 
P_z^{(+,0)}(\Lambda; x,y,z) &=& \frac
{Q'(x)}{Q(x)} \> 
\frac{\Delta D_{\Lambda/s}(z)}{D_{\Lambda/q}(z)} \> \hat A_{LL}(y)
\label{sL0L} \\ 
P_z^{(+,+)}(\Lambda; x,y,z) &=& \frac
{Q'(x) \, \hat A_{LL}(y) + \Delta Q'(x)}
{Q(x) + \Delta Q(x) \, \hat A_{LL}(y)} \> 
\frac{\Delta D_{\Lambda/s}(z)}{D_{\Lambda/q}(z)} \label{sL-SLL} \\ 
P_z^{(+,-)}(\Lambda; x,y,z) &=& \frac
{Q'(x) \, \hat A_{LL}(y) - \Delta Q'(x)}
{Q(x) - \Delta Q(x) \, \hat A_{LL}(y)} \>
\frac{\Delta D_{\Lambda/s}(z)}{D_{\Lambda/q}(z)} \label{sL+SLL}  
\earr
where
\barr
Q &\equiv& 4(u + \bar u) + (d + \bar d) + (s + \bar s) 
\label{Qtot} \\
\Delta Q &\equiv& 4(\Delta u + \Delta \bar u) + (\Delta d + \Delta \bar d) 
+ (\Delta s + \Delta \bar s) 
\label{dQtot} \\
Q' &\equiv& [4(u + \bar u) + (d + \bar d)] \, N_u + (s + \bar s)
\label{Qp} \\
\Delta Q' &\equiv& [4(\Delta u + \Delta \bar u) + (\Delta d + \Delta \bar d)]
\, N_u + (\Delta s + \Delta \bar s) \>.
\label{dQp}
\earr

If one assumes the same relation (\ref{ddq}) to hold also for 
transversely polarized quark fragmentation functions, then Eq. (\ref{0SN})
yields:
\beq
P_y^{(0,S_N)}(\Lambda; x,y,z) = \frac
{\Delta_T Q'(x)}{Q(x)} \> 
\frac{\Delta_T D_{\Lambda/s}(z)}{D_{\Lambda/q}(z)} \> \hat D_{NN}(y)
\label{0SNT} 
\eeq 
with
\beq
\Delta_T Q' \equiv [4(\Delta_T u + \Delta_T \bar u) + 
(\Delta_T d + \Delta_T \bar d)] \, N_u + (\Delta_T s + \Delta_T \bar s) \>.
\label{dtQp}
\eeq

Eqs. (\ref{0-SLL})-(\ref{sL+SLL}) hold under assumptions (\ref{dl}) and 
(\ref{ddq}) alone, independently of the actual value of $N_u$; 
as the polarized and unpolarized distribution functions are 
well known and several sets are available in the literature, we can exploit 
Eqs. (\ref{0-SLL}) and (\ref{sL0L}) to obtain:
\beq
N_u = - \frac{S}{U} \> \frac {\hat A_{LL} \> P_z^{(0,+)} - 
(\Delta S/S) \> P_z^{(+,0)}} {\hat A_{LL} \> P_z^{(0,+)} - 
(\Delta U/U) \> P_z^{(+,0)}} \label{nu}
\eeq
and
\beq
\frac{\Delta D_{\Lambda/s}(z)}{D_{\Lambda/q}(z)} = 
\frac {Q}{S} \> \frac{1}{\hat A_{LL}} \> 
\frac {(\Delta U/U) \> P_z^{(+,0)} - \hat A_{LL} \> P_z^{(0,+)}}
{(\Delta U/U) - (\Delta S/S)} \label{ddd}
\eeq
where we have defined
\beq
U \equiv 4(u + \bar u) + (d + \bar d)  \quad S \equiv s + \bar s
\quad \Delta U \equiv 4(\Delta u + \Delta \bar u) + (\Delta d + \Delta \bar d)
\quad \Delta S \equiv \Delta s + \Delta \bar s \>. \label{US}
\eeq

Eqs. (\ref{sL-SLL}) and (\ref{sL+SLL}) can then be used to predict -- within
the general assumptions (\ref{dl}) and (\ref{ddq}) -- the following 
interesting relations between polarization observables:
\beq
P_z^{(+,\pm)} = \frac {P_z^{(+,0)} \pm P_z^{(0,+)}} {1 \pm (\Delta Q/Q) 
\> \hat A_{LL}} \label{pz++-}
\eeq
and
\beq
P_z^{(+,+)} - P_z^{(+,-)} = 2 \> \frac {P_z^{(0,+)} - (\Delta Q/Q) \> 
\hat A_{LL} \> P_z^{(+,0)}}
{1 - [(\Delta Q/Q) \> \hat A_{LL}]^2} \> \cdot \label{p++-p+-}
\eeq  

Notice that in the small $y$ region, due to the elementary
dynamics, see Eq. (\ref{all}), one has:
\beq P_z^{(+,0)} \simeq 0 \quad\quad\quad P_z^{(+,\pm)} \simeq \pm P_z^{(0,+)}
\quad\quad\quad (y \ll 1) \>. \label{smy}
\eeq

In the large $y$ region instead, again from Eq. (\ref{all}) and from the 
fact that large $y$ implies small $x$, where $Q(x) \gg \Delta Q(x)$ and
$Q'(x) \gg \Delta Q'(x)$, we expect:
\beq P_z^{(+,+)} \simeq P_z^{(+,-)} \simeq P_z^{(+,0)}
\quad\quad\quad (y \simeq 1) \>. \label{lgy}
\eeq

We conclude this Section by reminding that we have derived our results
for $\Lambda = \Lambda^0 + \bar\Lambda^0$, which allows assumption 
(\ref{dl}) concerning $q$ and $\bar q$ fragmentation functions.
However, Eqs. (\ref{0-SLL})-(\ref{sL+SLL}) and (\ref{0SNT}) hold identical 
also for single $\Lambda^0$ production, provided one neglects the 
fragmentation function of a $\bar q$ into $\Lambda^0$, {\it i.e.} one 
neglects all $\bar q$ terms in Eqs. (\ref{Qtot})-(\ref{dQp}) and (\ref{dtQp}).
Anyway, the production of $\bar\Lambda^0$, in $\ell p$ processes is 
strongly suppressed by the limited amount of initial $\bar q$, unless one
considers very small $x$ values.   
   
\vskip 18pt
\nd
{\bf 4. Numerical estimates}
\vskip 6pt

We give now some numerical estimates of Eqs. (\ref{0-SLL})-(\ref{sL+SLL}) 
and (\ref{0SNT}); we use the sets of unpolarized and polarized fragmentation 
functions introduced and discussed by the authors of Ref. \cite{vog}: 
together with Eqs. (\ref{dl}), we use the expression for the unpolarized 
fragmentation functions they obtained by fitting $e^+ e^- \to \Lambda X$ 
data. At initial $Q^2_0 = 0.23$ (GeV/$c)^2$ scale one has, from a leading 
order (LO) analysis \cite{vog}:
\beq
D_{\Lambda/q}(z, Q^2_0) = 0.63 \, z^{0.23} (1-z)^{1.83} \>. \label{dlin}
\eeq

The polarized fragmentation functions are assumed \cite{vog} to be of the 
initial form (\ref{ddq}), with:
\beq
\Delta D_{\Lambda/s}(z, Q^2_0) = z^\alpha D_{\Lambda/q}(z, Q^2_0) \>.
\label{dds}
\eeq
Leading order QCD evolution is consistently taken into account in our 
numerical computations.\footnote{We are very grateful to D. de Florian, 
M. Stratmann and W. Vogelsang for providing us with their FORTRAN package 
for unpolarized and polarized fragmentation functions.} Next to leading 
order contributions to the $\Lambda$ 
polarization have been shown to be tiny \cite{vog} and we neglect them. 
 
The parameter $N_u$ defined in Eq. (\ref{ddq}), 
has been chosen according to three different scenarios typical of a wide 
range of plausible models, and the corresponding remaining 
parameter $\alpha$ of Eq. (\ref{dds}) has been fixed by fitting the few 
LEP data on $\Lambda$ polarization, with the results \cite{vog}:

\begin{enumerate}
\item[1)]  
{\mbox{\boldmath $N_u = 0, \> \alpha = 0.62$}}.
This scenario corresponds to $SU(6)$ non relativistic quark model, 
according to which the whole $\Lambda$ spin is carried by the $s$ quark.

\item[2)]  
{\mbox{\boldmath $N_u = -0.2, \> \alpha = 0.27$}}.
Such value of $N_u$ is suggested in Ref. \cite{bj}, based on a $SU(3)$ 
flavour symmetry analysis and on data on the first moment of $g_1^p$.

\item[3)]
{\mbox{\boldmath $N_u = 1, \> \alpha = 1.66$}}.
A scenario in which, contrary to the non relativistic models, all 
light quarks contribute equally to the $\Lambda$ polarization. 

\end{enumerate}

The unpolarized and polarized distribution functions are taken respectively 
from Refs. \cite{grv} and \cite{dq}. We have explicitely checked that a 
different choice of unpolarized and polarized distribution functions,
like those of Refs. \cite{lss1} and \cite{lss2}, can change significantly
the numerical values of $P_i(\Lambda)$ (but not their qualitative
behaviour) only for $x \gsim 0.3$.

A computation of $P_y^{(0,S_N)}$, Eq. (\ref{0SNT}), requires the knowledge
of the quark transversely polarized distributions, $\Delta_Tq$ or $h_{1q}$,
and of the transversely polarized fragmentation functions, $\Delta_T D$,
which are not known. In order to give an estimate we fix $\Delta_Tq$
for $u$ and $d$ quarks by saturating the Soffer's bound \cite{sof}
(assuming the same signs for $\Delta_Tq$ and $\Delta q$):
\beq
\Delta_Tu = \frac{1}{2}\>(u + \Delta u) \quad\quad    
\Delta_Td = -\frac{1}{2}\>(d + \Delta d) \>. \label{dtq}
\eeq
All other transverse distributions ($\Delta_T \bar q$ and $\Delta_T s$) are
neglected here and we also assume $\Delta_T D_{\Lambda/s} = 
\Delta D_{\Lambda/s}$.  

A sample of typical results is presented in Figs. 1-4, for HERMES
kinematics, $s = 52.4$ (GeV)$^2$ and $Q^2 \gsim 1$ (GeV/$c)^2$.

\begin{itemize}
\item
In Fig. 1a we plot $P_z^{(0,+)}$ -- at fixed $x=0.1$ and $z=0.5$ 
values -- as a function of $y$, for each of the three scenarios; $P_z^{(0,+)}$,
Eq. (\ref{0-SLL}), can depend on $y$ only via the $Q^2$-evolution and indeed 
the three curves show an almost flat behaviour. The three scenarios 
yield quite different results. The minimum value of $y$ is given by
$y_{min} = Q^2_{min}/(xs) \simeq 0.19$. 

The same plot for $P_z^{(+,0)}$ is presented in Fig. 1b; the $y$-dependence
is essentially due to the factor $\hat A_{LL}$ in Eq. (\ref{sL0L}); scenario
3, which assumes $Q'(x) = Q(x)$, together with Eq. (\ref{dds}) in which
we neglect the mild $Q^2$-evolution (taken into account in our numerical
computations), gives a particularly simple result:
\beq
P_z^{(+,0)} \simeq z^{1.66} \, \hat A_{LL} \>. \label{0+3}
\eeq

\item
$P_z^{(+,+)}$ and $P_z^{(+,-)}$, again as functions of $y$ at fixed 
$x=0.1$ and $z=0.5$ values, are shown in Figs. 2a and 2b respectively.
At large $y \to 1$ values Eq. (\ref{lgy}) is satisfied. The small 
$y \to 0$ behaviour cannot be seen with $x=0.1$; in Fig. 2c and 2d we plot
respectively $P_z^{(+,+)}$ and $P_z^{(+,-)}$, changing the $x$ value to 
$x=0.3$ and keeping $z=0.5$. The allowed minimum value of $y$ is now 0.06
and we have checked that Eq. (\ref{smy}) is indeed obeyed (by comparing 
with $P_z^{(0,+)}$ as a function of $y$ at $x=0.3$ and $z=0.5$, not shown 
in Fig. 1); the change in sign of $P_z^{(+,-)}$ is particularly interesting.  

\item   
In Figs. 3a-3d we show respectively $P_z^{(0,+)}$, $P_z^{(+,0)}$,
$P_z^{(+,+)}$ and $P_z^{(+,-)}$, at fixed values of $Q^2 = 1.7$ (GeV/$c)^2$
and $z=0.5$, as functions of $x$. At fixed $Q^2$, $y$ decreases with 
increasing $x$ -- and viceversa -- and this explains why relations 
(\ref{smy}) and (\ref{lgy}) hold respectively when $x \to 1$ and $x \to 0$.  
Once more, the three different scenarios give very different 
results. Using different sets of polarized distribution functions, 
like \cite{lss1} or \cite{lss2} instead of \cite{dq}, gives almost 
identical results for $x \lsim 0.3$ and larger (in magnitude) results
for $x \gsim 0.3$ (of course, $P_z^{(+,0)}$ is not affected at all by a change 
in $\Delta q$). 

\item
In Figs. 4a we plot $P_y^{(0,S_N)}$, Eq. (\ref{0SNT}), at fixed 
$x=0.1$ and $z=0.5$ values, as a function of $y$ for all scenarios;
the $y$ dependence is almost entirely given by $\hat D_{NN}$, Eq. (\ref{dnn}),
and indeed  $P_y^{(0,S_N)} \to 0$ when $y \to 1$. Notice the opposite
$y$ behaviour of $P_z^{(+,0)}$ and $P_y^{(0,S_N)}$ due to the opposite
behavior of $\hat A_{LL}$ and $\hat D_{NN}$.

In Fig. 4b $P_y^{(0,S_N)}$ is plotted as a function of $x$ at fixed  
$Q^2 = 1.7$ (GeV/$c)^2$ and $z=0.5$. 

\item
In general one finds very small values of $P_i(\Lambda)$ in scenario 1,
negative values in scenario 2 and positive ones in scenario 3. This can 
easily be understood by the different values of $N_u$ in the three 
scenarios, which assign respectively zero, negative, and positive 
contributions to $u$ and $d$ quarks, which dominate in the proton.
Experimental measurements can easily discriminate between them. 

\item
Recently HERMES Collaboration \cite{her} have published a single experimental
measurement of $P_z^{(+,0)}/\hat A_{LL}$, as a function of $z$ and this
seems to favour the scenario 1 prediction of Ref. \cite{vog}, although
errors and uncertainties are still too large to draw any reliable conclusions.
Similarly, two values of $P_z^{(+,0)}(z)$ published by the E665 Collaboration 
\cite{e665} still have much too large errors.  

\end{itemize}

\vskip 18pt
\nd
{\bf 5. Conclusions}
\vskip 6pt
The study of the angular distribution of the $\Lambda \to p\pi$
decay allows a simple and direct measurement of the components
of the $\Lambda$ polarization vector. For $\Lambda$'s produced
in the current fragmentation region in DIS processes, the component 
of the polarization vector are related to spin properties of the
quark inside the nucleon, to spin properties of the quark hadronization, 
and to spin dynamics of the elementary interactions.
All this information, concerning quark distribution functions, 
quark fragmentation functions and spin properties of elementary dynamics
are essentially factorized and separated as depending on three different
variables, respectively $x$, $z$ and $y$. The $Q^2$-evolution and 
dependence of distribution and fragmentation functions somewhat mix
the three variables, but smoothly, keeping separated the main properties 
of each of the different aspects of the process. Moreover, such $Q^2$
dependence is perturbatively well known and under control.   
  
We have discussed all different polarization states of baryons, obtainable
in the fragmentation of a quark in DIS with polarized initial leptons 
and nucleons, Eqs. (\ref{0-SL})-(\ref{sL+SL}), showing how they can reveal
different quark features, weighted and shaped by elementary dynamics.

Adopting a simplifying -- although rather general and representative of
many possible choices -- assumption about the quark fragmentation functions
into a $\Lambda$ \cite{vog}, we are able to extract from measurements further 
information on the quark fragmentation process, Eqs. (\ref{nu}) and 
(\ref{ddd}), and to predict relations among polarization states induced by 
different initial spin configurations, Eqs. (\ref{pz++-}) and (\ref{p++-p+-}).
 
Numerical estimates are given in Figs. 1-4, according to three largely
different scenarios \cite{vog} for fragmentation functions; each scenario
has physical motivations and yields qualitatively different results: 
compatible with zero, large and negative, large and positive. 
Such results are stable against different choices of the polarized and 
unpolarized distribution functions, so that experimental information should
immediately allow to draw clear conclusions and to learn about quark
fragmentation properties.      

The elementary dynamics fixes the small or large $x$ or $y$ behaviour
of some of the polarization components; although expected, such behaviours
should indeed be checked, as an independent and non trivial test of the
QCD factorization scheme of Eq. (\ref{gen}); such a scheme has been
widely used and tested for the computation of semi-inclusive unpolarized 
cross-sections, but not for more subtle spin observables. 

We think that our comparative and comprehensive discussion 
of all possible $\Lambda$ polarization measurements in polarized DIS is 
useful and can lead to a new and clear strategy which allows to obtain 
novel information.

\vskip 18pt 
\nd  
{\bf Acknowledgements}
\vskip 6pt
We would like to thank W. Vogelsang for useful discussions.


\newpage

\begin{figure}[t]
\[\begin{array}{ll} \hspace*{-1.0cm}
\mbox{~\epsfig{file=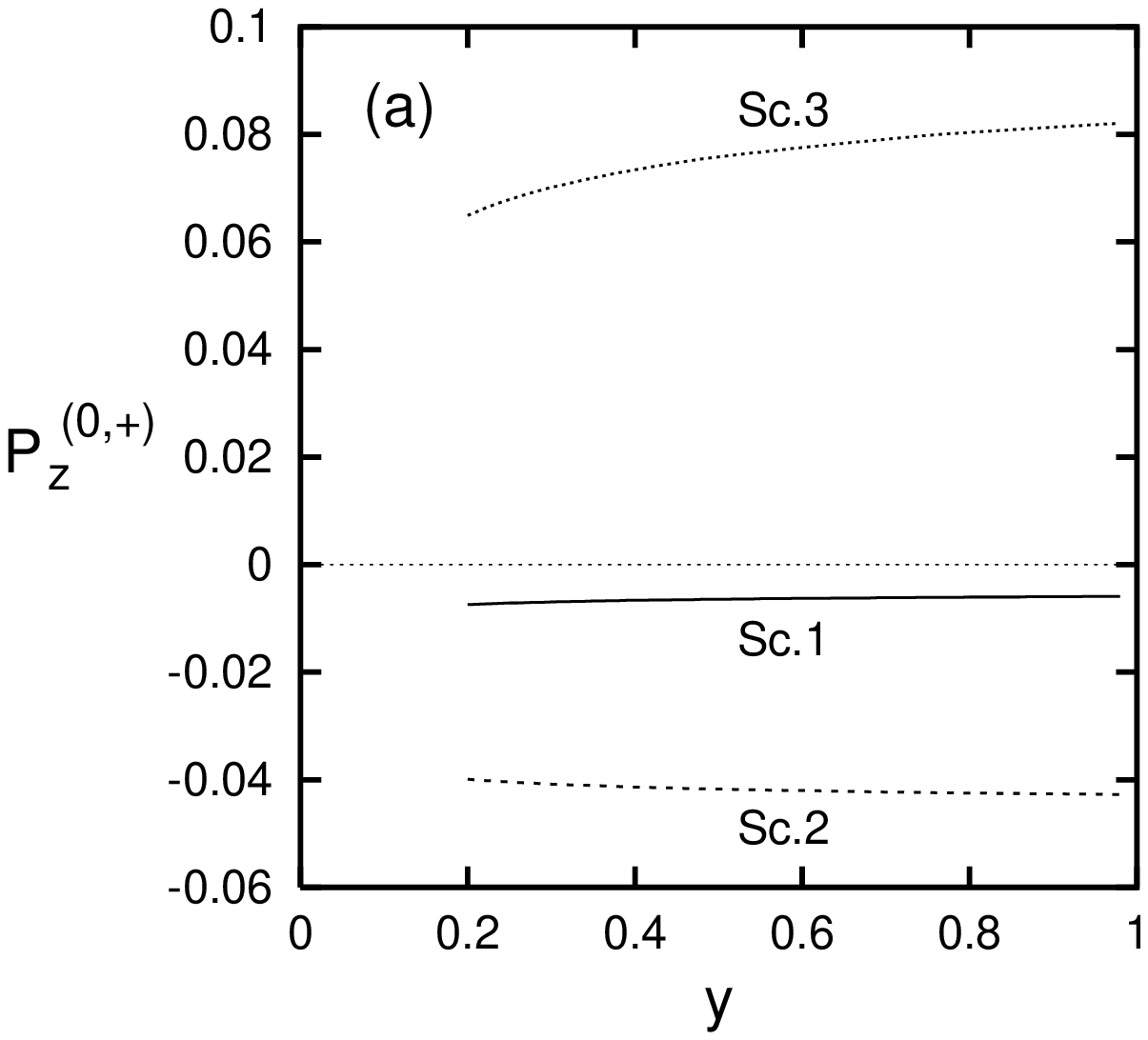,angle=-90,width=7.6cm}} 
& 
\mbox{~\epsfig{file=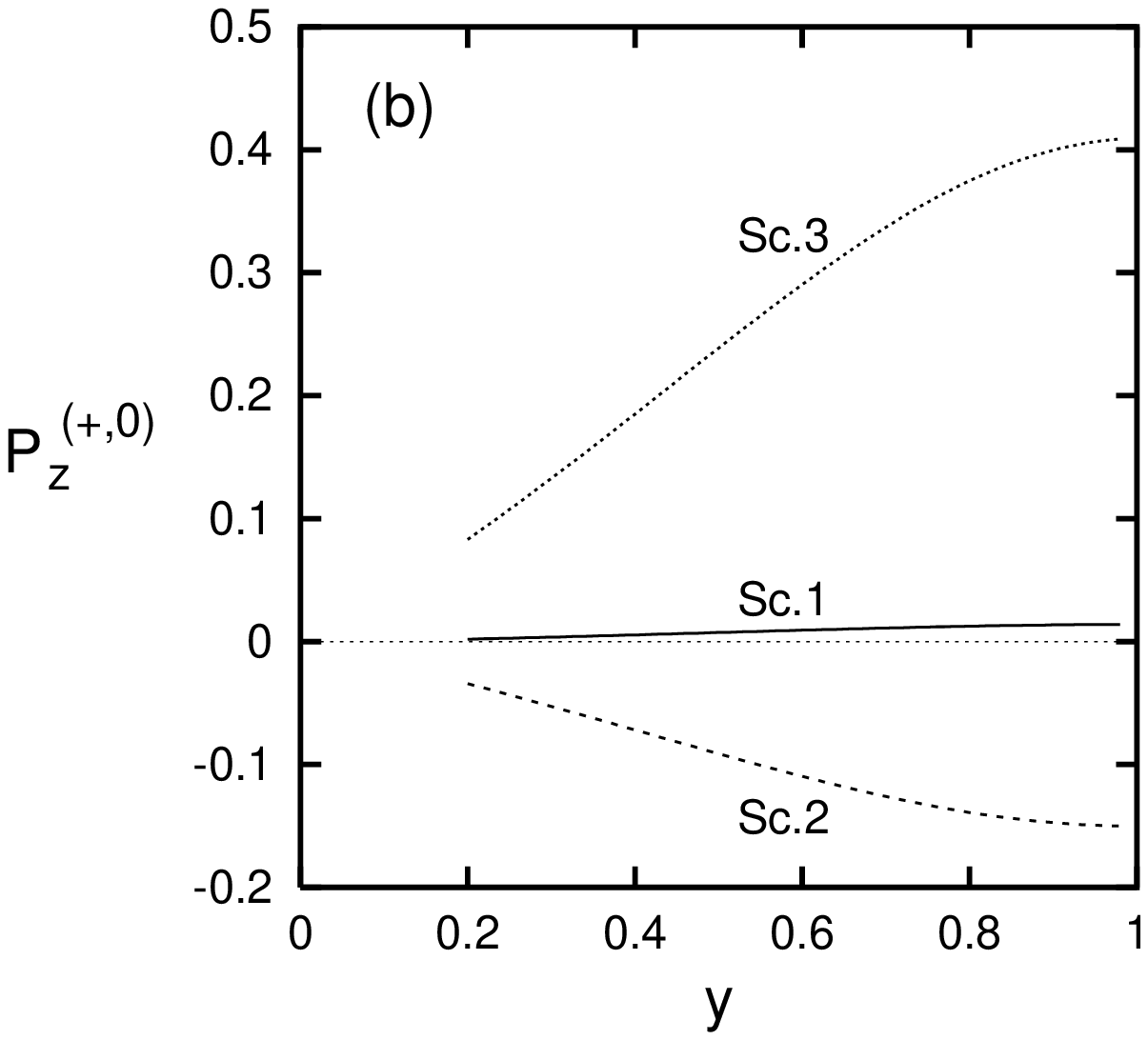,angle=-90,width=7.6cm}} 
\end{array}\]
\vspace{0.6cm}
\caption{\label{fig1}
$P_z^{(0,+)}$ and $P_z^{(+,0)}$ as a function of $y$ at fixed values of 
$x=0.1$ and $z=0.5$, for the three different scenarios.}
\end{figure}

\newpage

\begin{figure}[t]
\[\begin{array}{ll} \hspace*{-1.0cm}
\mbox{~\epsfig{file=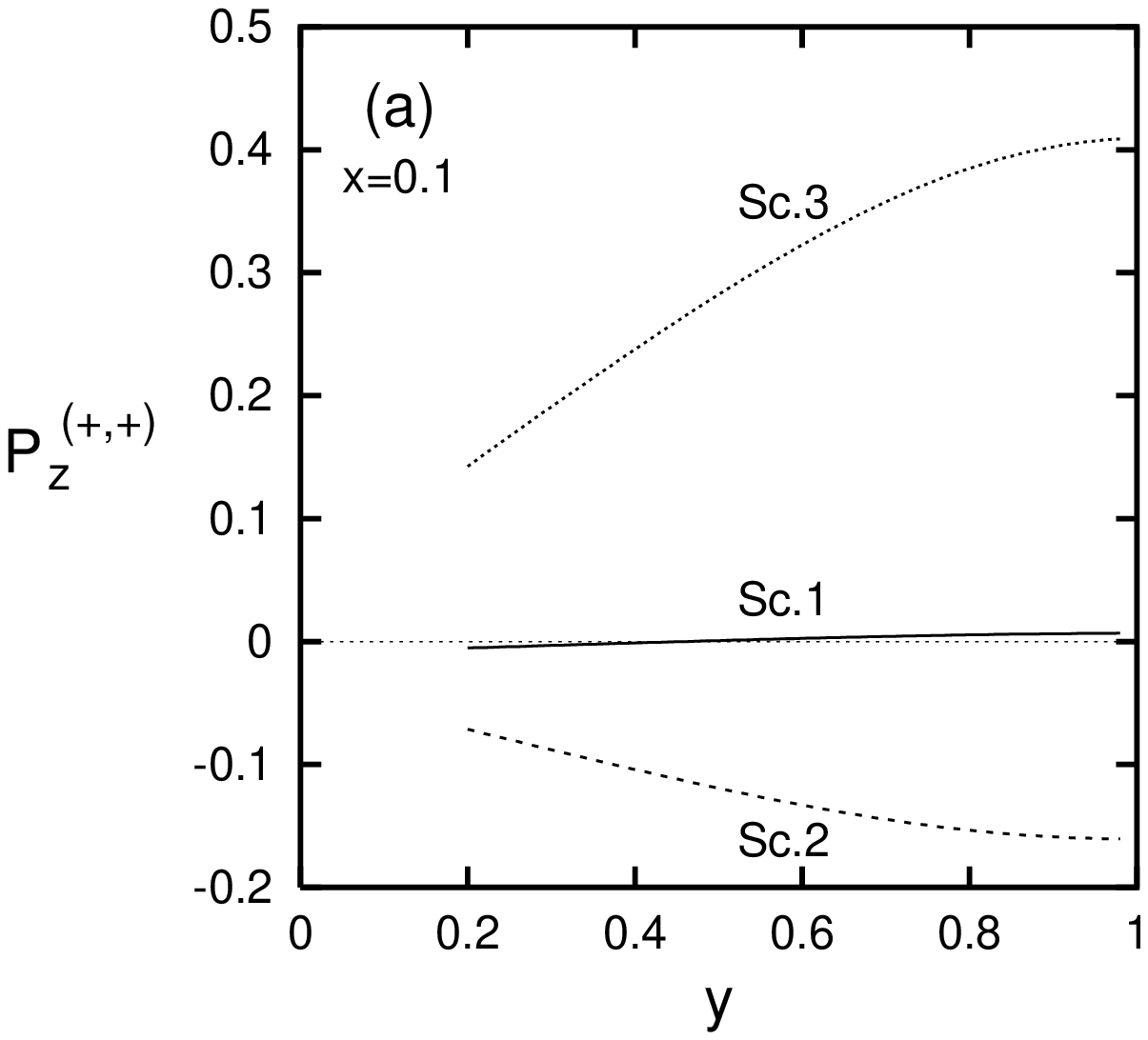,angle=-90,width=7.6cm}} 
& 
\mbox{~\epsfig{file=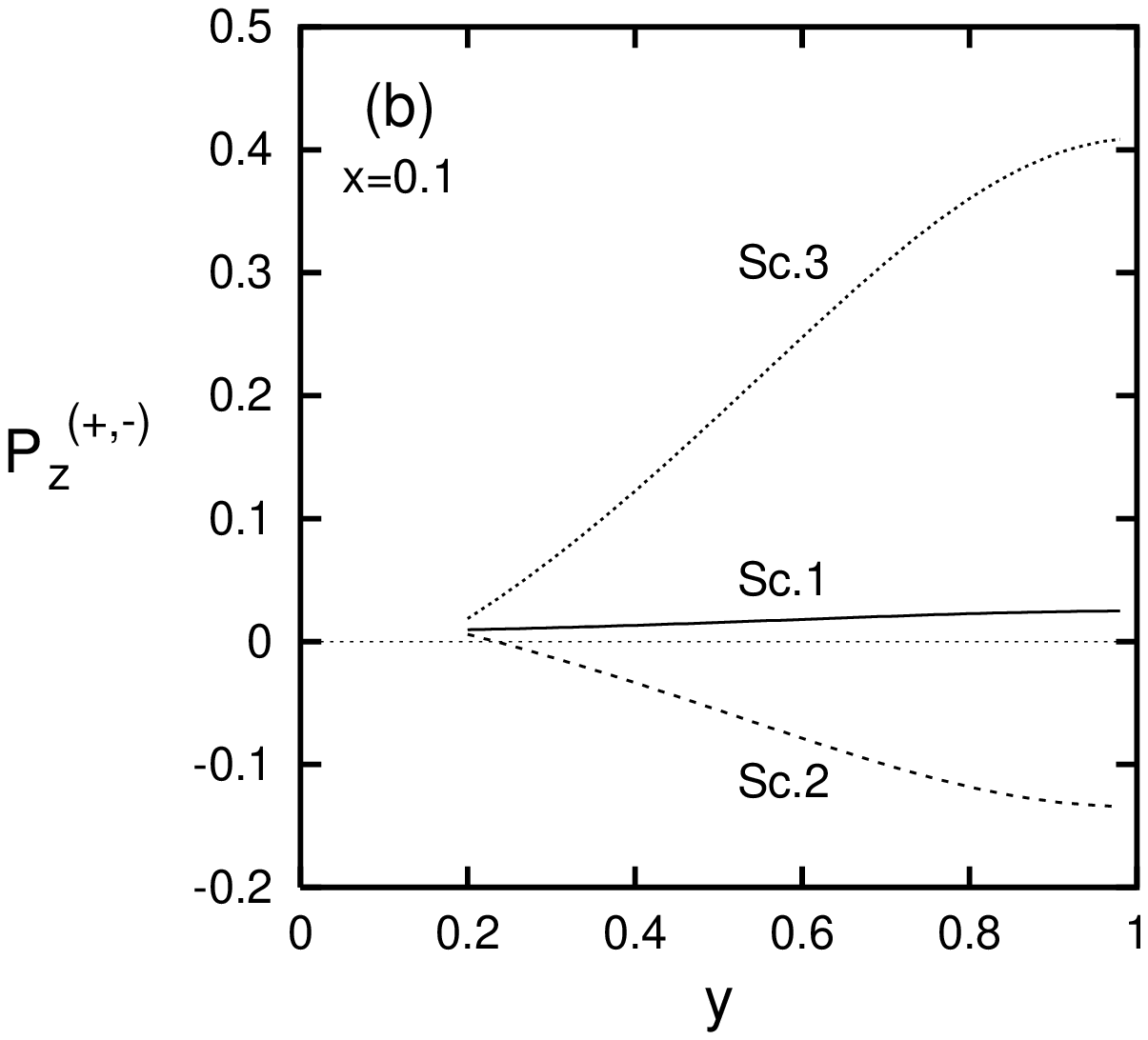,angle=-90,width=7.6cm}} 
\\ \hspace*{-1.0cm}
\mbox{~\epsfig{file=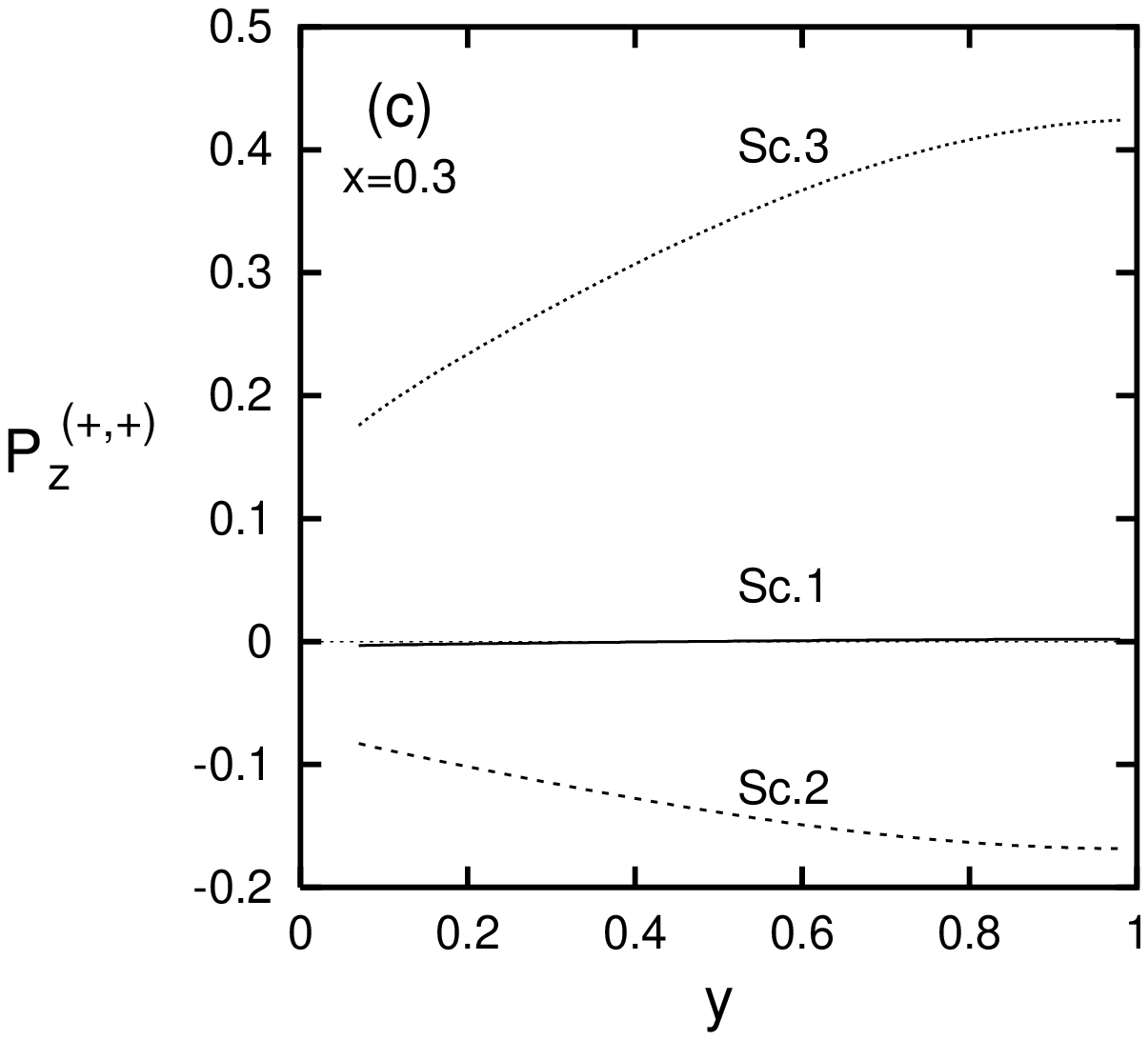,angle=-90,width=7.6cm}} 
& 
\mbox{~\epsfig{file=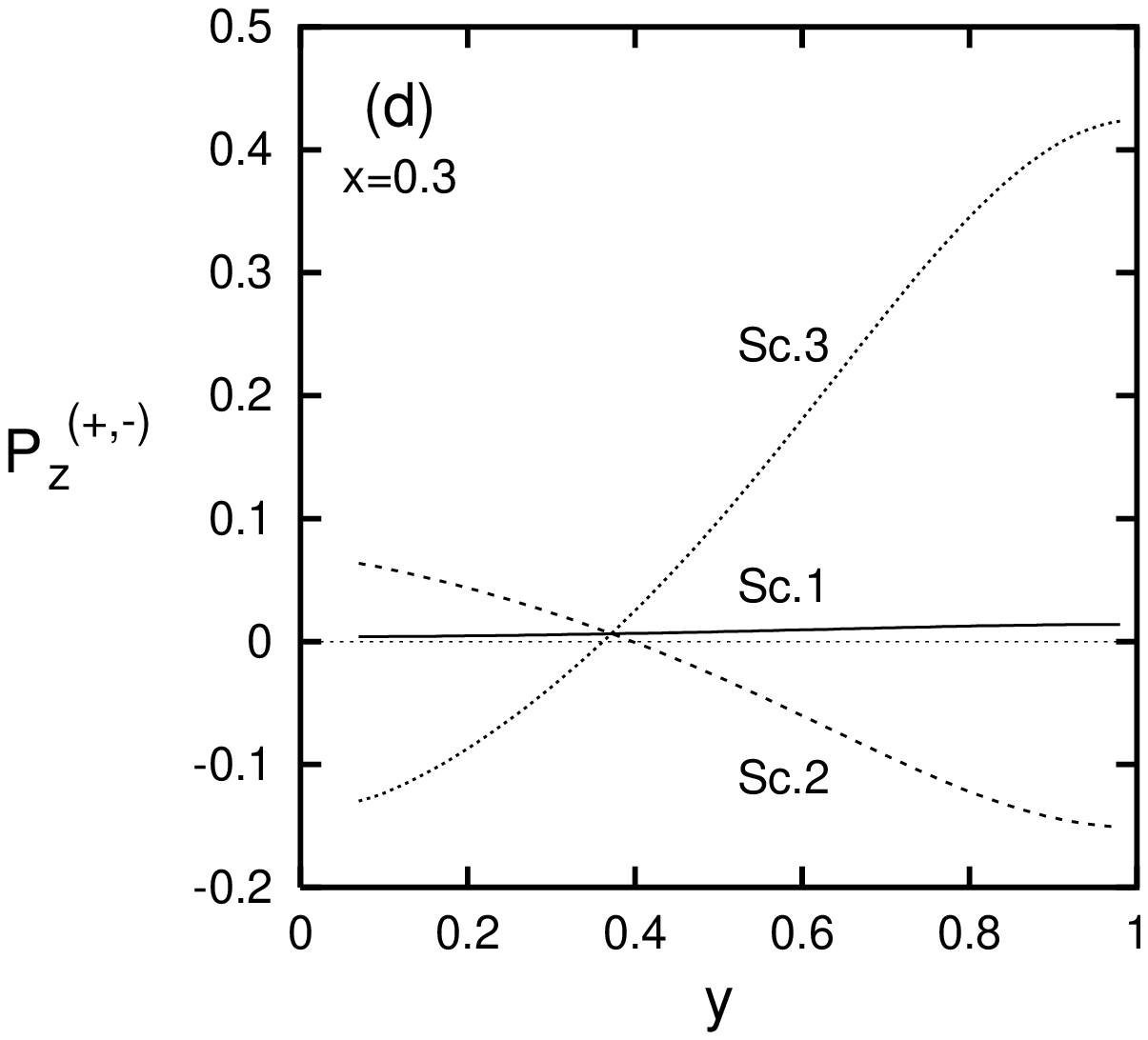,angle=-90,width=7.6cm}} 
\end{array}\]
\vspace{0.6cm}
\caption{\label{fig2}
$P_z^{(+,+)}$ and $P_z^{(+,-)}$ as a function of $y$ at fixed values of 
$z=0.5$ for the three different scenarios. The upper plots correspond 
to fixed $x=0.1$ and the lower plots to $x=0.3$.}
\end{figure}

\newpage

\begin{figure}[t]
\[\begin{array}{ll} \hspace*{-1.0cm}
\mbox{~\epsfig{file=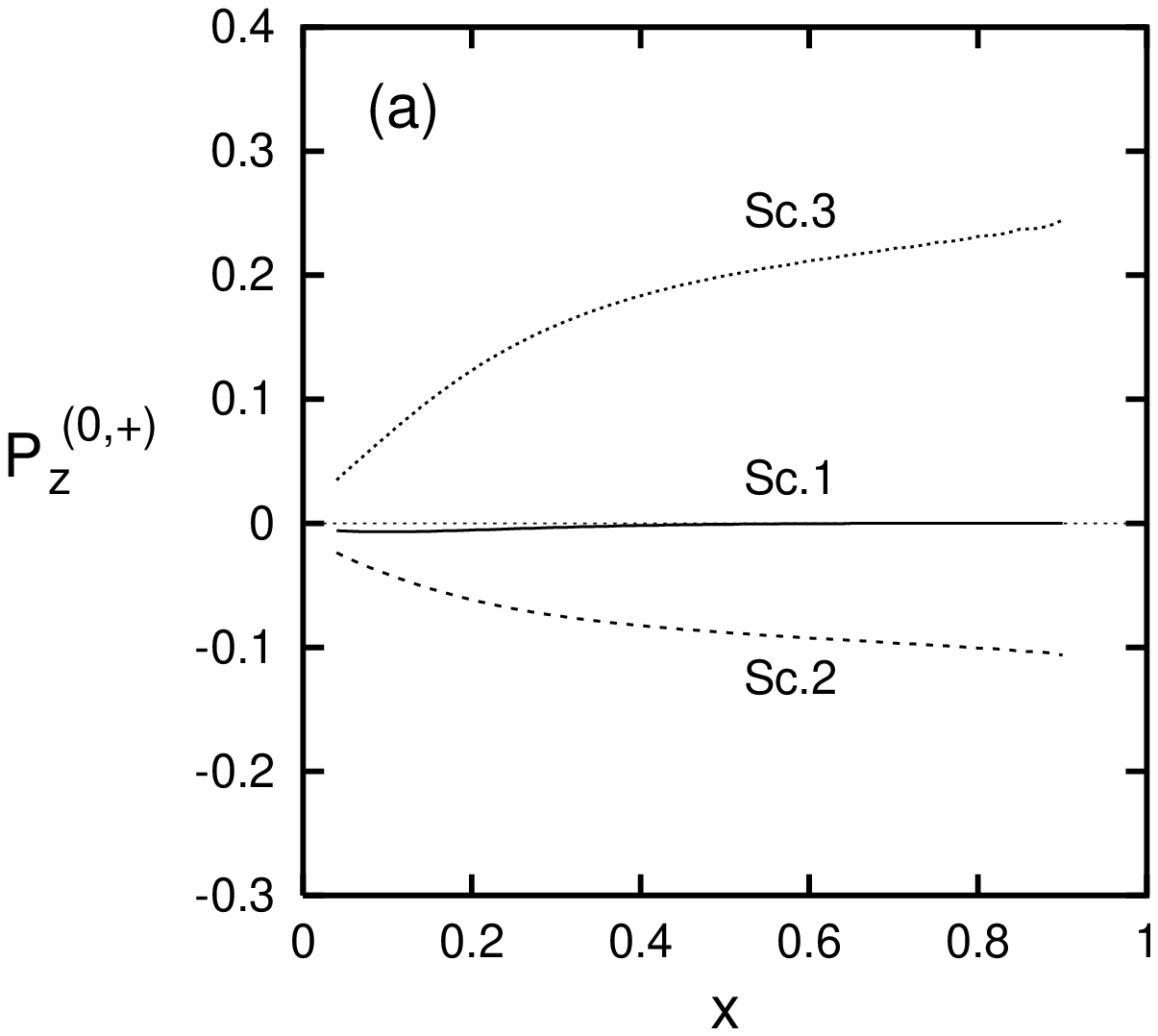,angle=-90,width=7.6cm}} 
& 
\mbox{~\epsfig{file=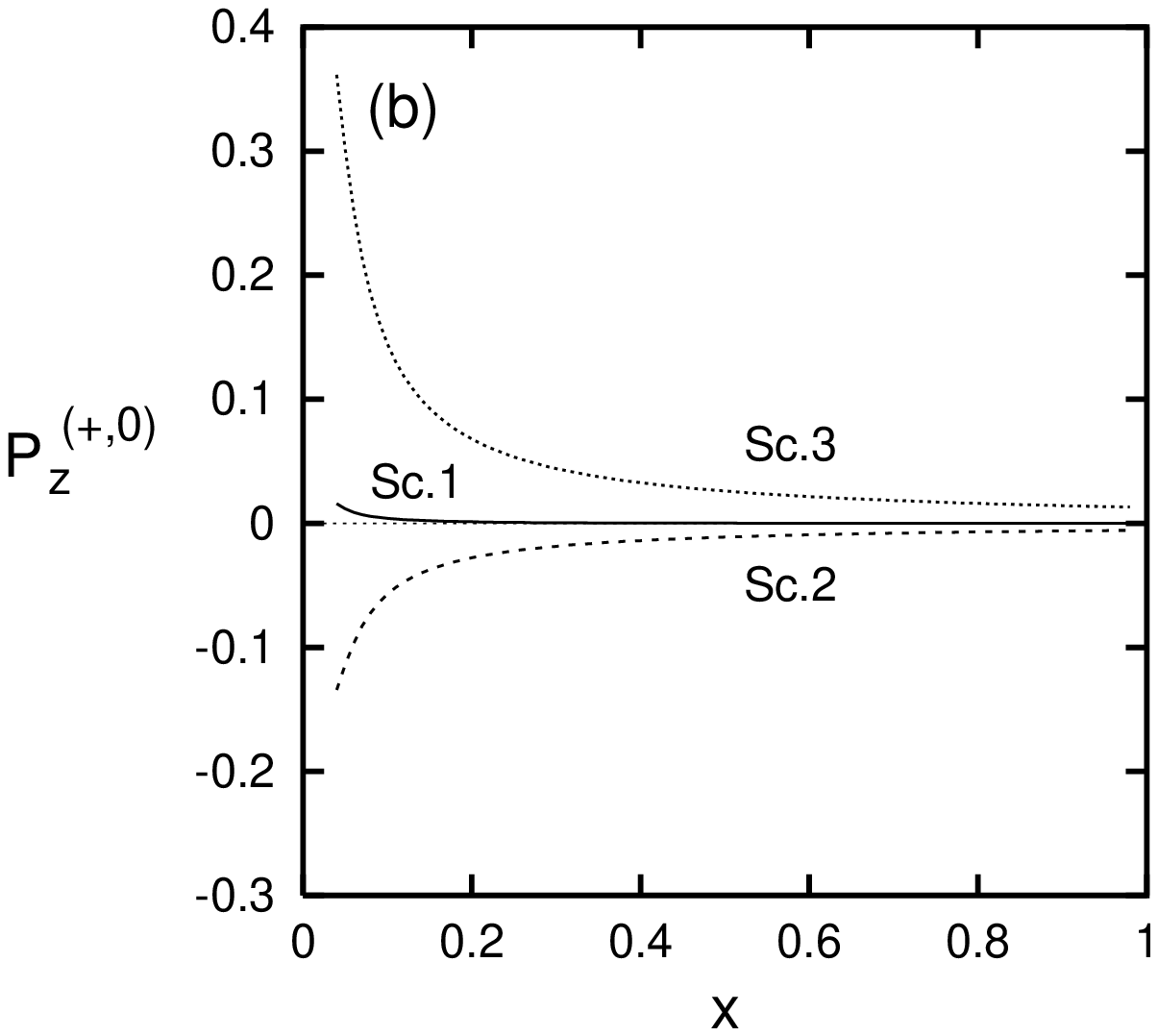,angle=-90,width=7.6cm}} 
\\ \hspace*{-1.0cm}
\mbox{~\epsfig{file=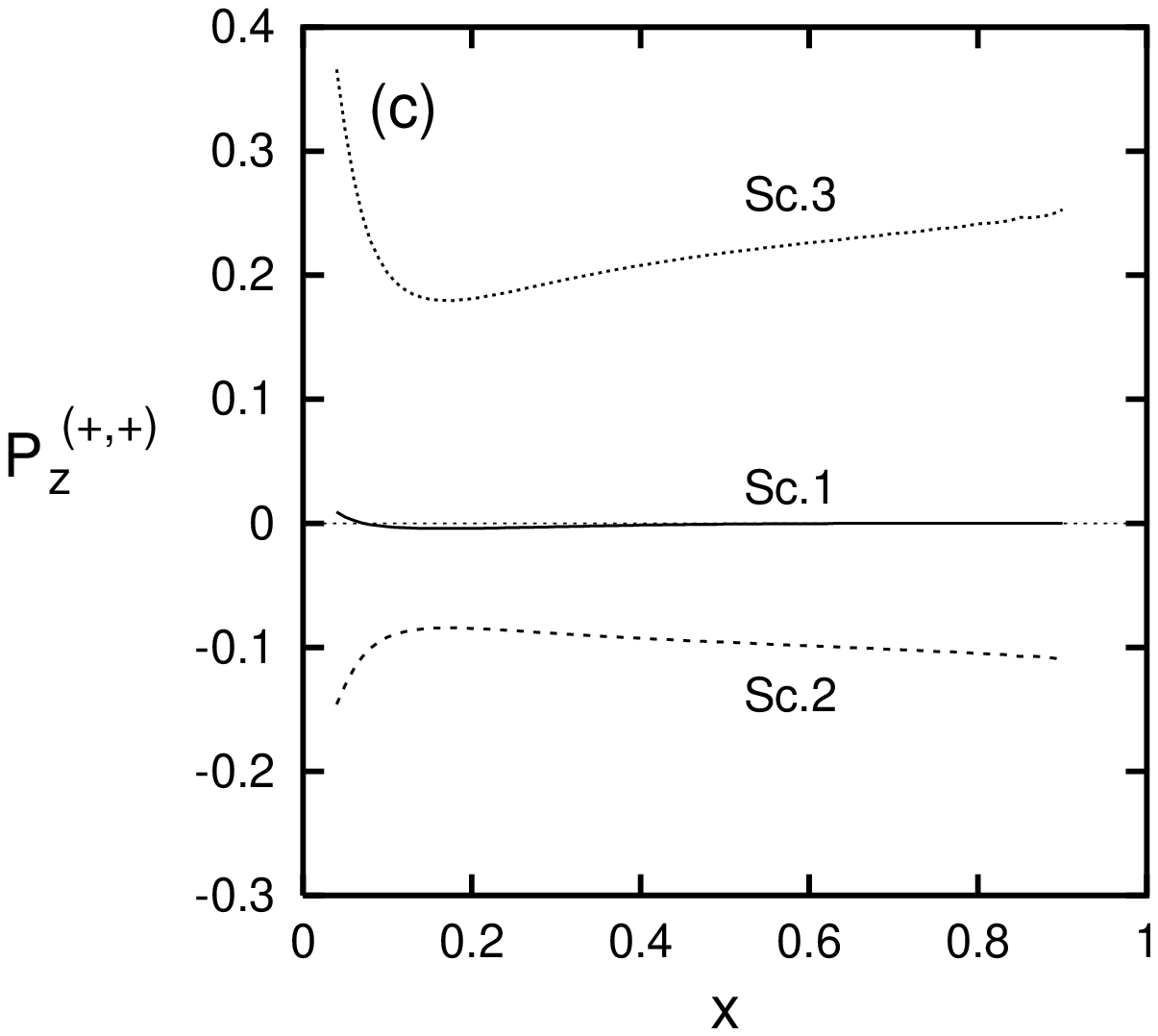,angle=-90,width=7.6cm}} 
& 
\mbox{~\epsfig{file=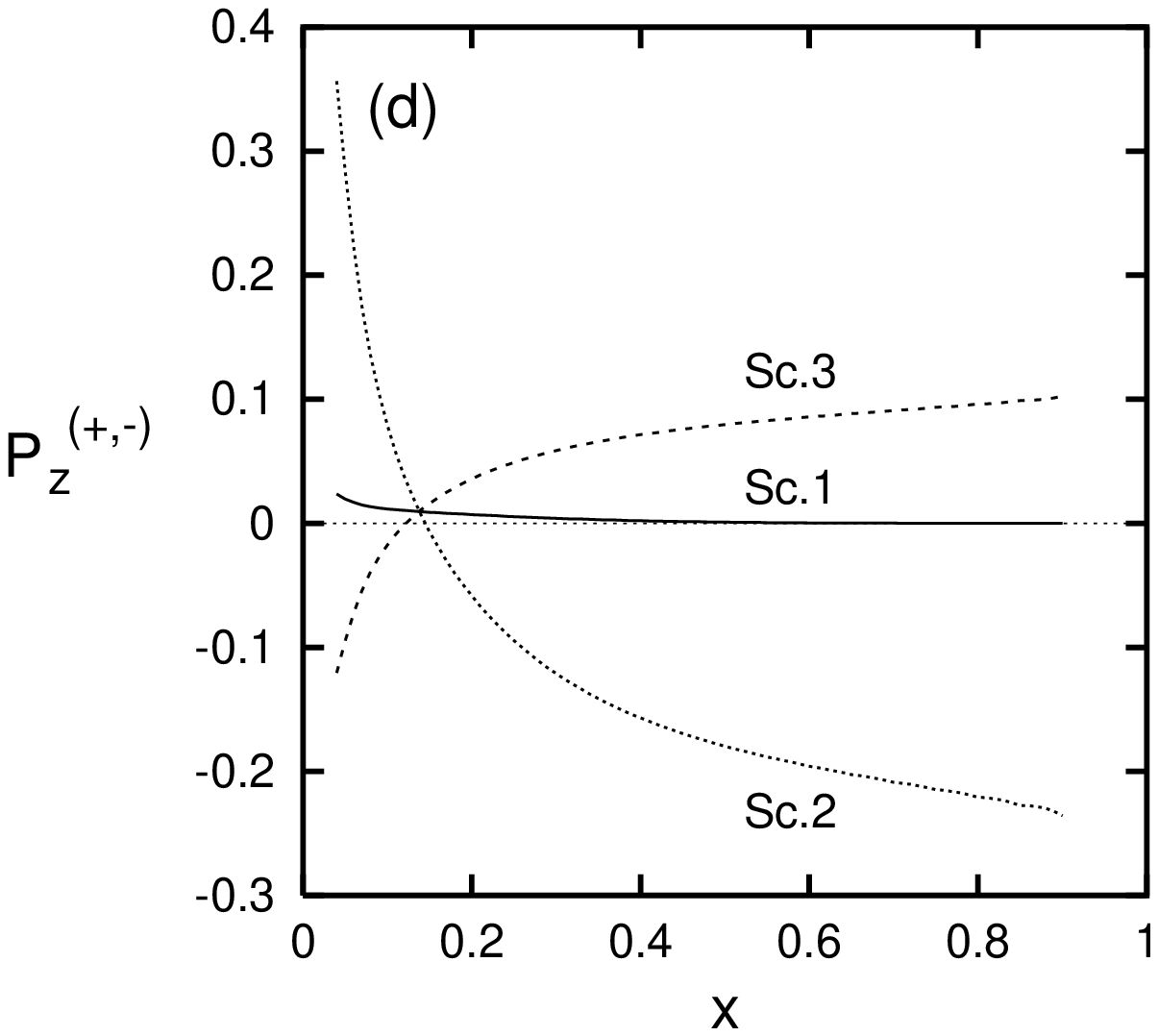,angle=-90,width=7.6cm}} 
\end{array}\]
\vspace{0.6cm}
\caption{\label{fig3}
$P_z^{(0,+)}$, $P_z^{(+,0)}$, $P_z^{(+,+)}$ and $P_z^{(+,-)}$ as a function 
of $x$ at fixed values of $Q^2=1.7$ (GeV/c)$^2$ and  $z=0.5$, for the three 
different scenarios.
}
\end{figure}

\newpage

\begin{figure}[t]
\[\begin{array}{ll} \hspace*{-1.0cm}
\mbox{~\epsfig{file=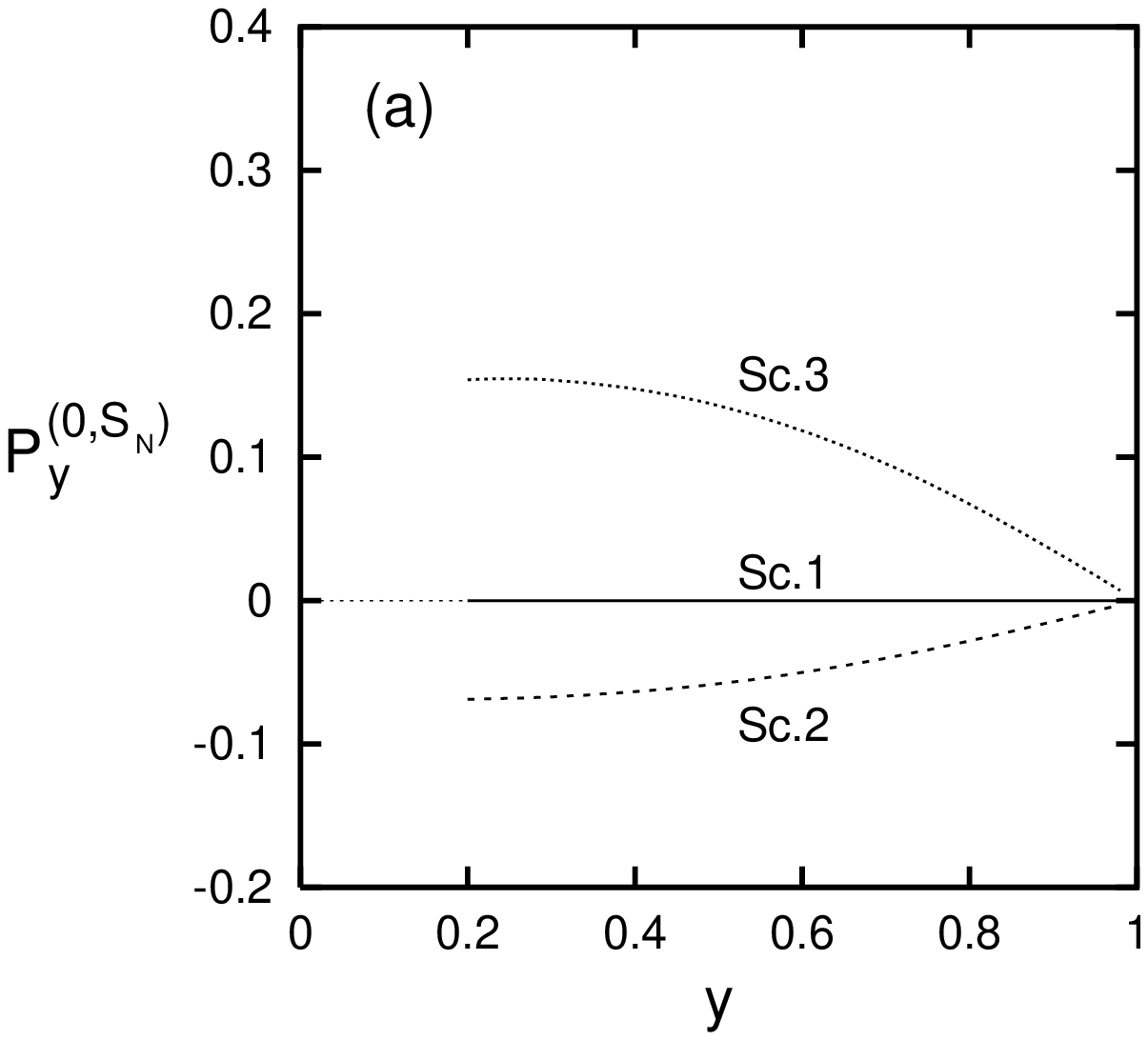,angle=-90,width=7.6cm}} 
& 
\mbox{~\epsfig{file=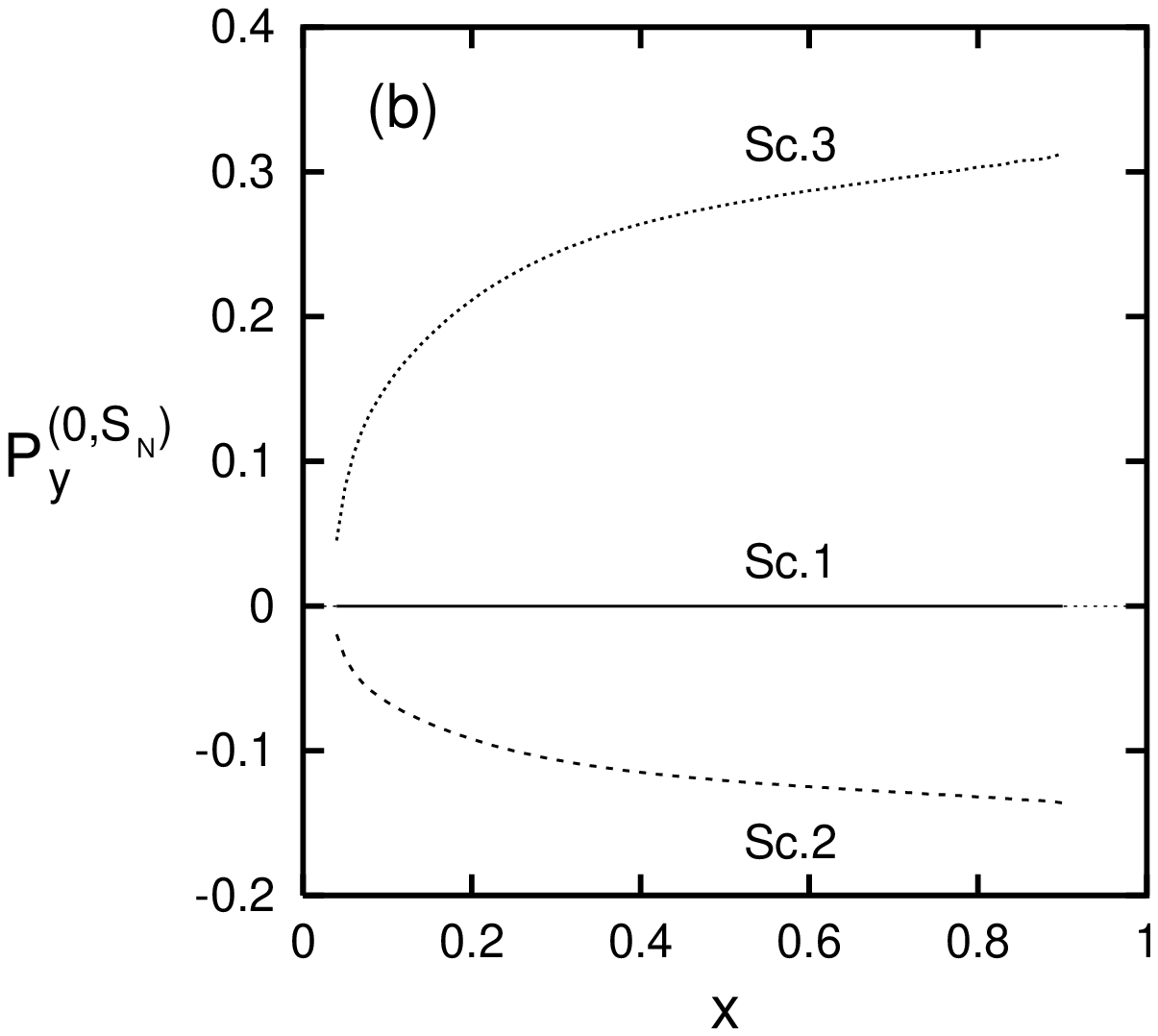,angle=-90,width=7.6cm}} 
\end{array}\]
\vspace{0.6cm}
\caption{\label{fig4}
(a), $P_y^{(0,S_N)}$ as a function of $y$ at fixed $x=0.1$ 
and $z=0.5$; \protect \\ (b), $P_y^{(0,S_N)}$  as a function of $x$ at 
fixed $Q^2=1.7$ (GeV/c)$^2$ and $z=0.5$.
}
\end{figure}


\newpage
\baselineskip=6pt
\small
\begin{thebibliography}{99}
\setlength{\parskip}{6pt}
\bibitem{jaf}
\vskip-8pt
R.L. Jaffe, \PR{D54} (1996) 6581
\bibitem{ekk}
\vskip-8pt
J. Ellis, D. Kharzeev and A. Kotzinian, \ZP{C69} (1996) 467
\bibitem{vog}
\vskip-8pt
D. de Florian, M. Stratmann and W. Vogelsang, \PR{D57} (1998) 5811 
\bibitem{kbv}
\vskip-8pt
A. Kotzinian, A. Bravar and D. von Harrach,
{\it Eur. Phys. J.} {\bf C2} (1998) 329
\bibitem{kot}
\vskip-8pt
A. Kotzinian, talk at the SPIN-97 VII Workshop on {\it High Energy
Spin Physics}, 7-12 July 1997, Dubna, Russia;
e-Print Archive: hep-ph/9709259
\bibitem{bel}
\vskip-8pt
S.L. Belostotski, \NP{B79} (Proc. Suppl.) (1999) 526
\bibitem{blt}
\vskip-8pt
C. Boros, J.T. Londergan and A.W. Thomas, e-Print Archive: hep-ph/9908260
\bibitem{al}
\vskip-8pt
D. Ashery and H.J. Lipkin, e-Print Archive: hep-ph/9908355 
\bibitem{mssy}
\vskip-8pt
B-Q. Ma, I. Schmidt, J. Soffer and J-Y. Yang, e-Print Archive: hep-ph/0001259 
\bibitem{noi}
\vskip-8pt
M. Anselmino, M. Boglione, J. Hansson and F. Murgia, \PR{D54} (1996) 828
\bibitem{am}
\vskip-8pt
X. Artru and M. Mekhfi, \ZP{C45} (1990) 669
\bibitem{bj}
\vskip-8pt
M. Burkardt and R.L. Jaffe, \PRL{70} (1993) 2537
\bibitem{grv}
\vskip-8pt
M. Gl\"uck, E. Reya and W. Vogelsang, \ZP{C67} (1995) 433
\bibitem{dq}
\vskip-8pt
M. Gl\"uck, E. Reya M. Stratmann and W. Vogelsang, \PR{D53} (1996) 4775
\bibitem{lss1}
\vskip-8pt
E. Leader, A.V. Sidorov and D.B. Stamenov, \PR{D58} (1998) 114028;
\PL{B462} (1999) 189
\bibitem{lss2}
\vskip-8pt
E. Leader, A.V. Sidorov and D.B. Stamenov, \IJMP{A13} (1998) 5573
\bibitem{sof}
\vskip-8pt
J. Soffer, \PRL{74} (1995) 1292
\bibitem{her}
\vskip-8pt
The HERMES Collaboration, A. Airapetian {\it et al.}, 
e-Print Archive: hep-ex/9911017
\bibitem{e665}
\vskip-8pt
The E665 Collaboration, M.R. Adams {\it et al.}, 
e-Print Archive: hep-ex/9911004

\end{thebibliography}
\end{document}